\documentclass[journal]{IEEEtran}

\pdfoutput=1
\usepackage{graphicx}
\ifCLASSINFOpdf
  
\else
\fi

\usepackage[cmex10]{amsmath}

\begin{document}

\title{Thermoelectrically Controlled Spin-Switch}

\author{\IEEEauthorblockN{S. Andersson and V. Korenivski} \\

	\IEEEauthorblockA{Nanostructure Physics, Royal Institute of Technology, SE-106 91 Stockholm, Sweden}}

\maketitle

\begin{abstract}
The search for novel spintronic devices brings about new ways to control switching in magnetic thin-films. In this work we experimentally demonstrate a device based on thermoelectrically controlled exchange coupling. The read out signal from a giant magnetoresistance element is controlled by exchange coupling through a weakly ferromagnetic Ni-Cu alloy. This exchange coupling is shown to vary strongly with changes in temperature, and both internal Joule heating and external heating is used to demonstrate magnetic switching. The device shows no degradation upon thermal cycling. Ways to further optimize the device performance are discussed. Our experimental results show a new way to thermoelectrically control magnetic switching in multilayers.

\end{abstract}


\section{Introduction}
\IEEEPARstart{C}{ontrolling} and optimizing the switching of thin magnetic films in a variety of spintronic devices has been of great interest since the discovery of giant magnetoresistance (GMR) \cite{ref:Binasch}-\cite{ref:Baibich}. Many ways of control have been demonstrated such as spin-torque \cite{ref:Katine}, toggle \cite{ref:Engel} and heat assisted switching \cite{ref:Beech} to mention a few. Today, the search for new applications has pushed the need even further. In thermally assisted MRAM \cite{ref:Prejbeanu} the information is stored by exchange coupling between an antiferromagnetic (AFM) and a ferromagnetic (FM) film. Switching of the FM film is controlled by heating the two films above the N\'eel temperature of the AFM and then cooling down in the presence of an external magnetic field. 

Recently a new device based on a novel way to control switching was proposed. By exchange coupling two FM films through a weakly FM alloy the coupling is controlled by changes in temperature. At room temperature the alloy is weakly FM and the two films are exchange coupled through the alloy. At a temperature higher than the Curie temperature the alloy is paramagnetic and the two strongly FM films decouple. Provided the films have separate switching fields their moments can then be aligned either parallel or antiparallel. Either external or internal heating can be used. By using a weakly FM alloy of Ni and Cu, we demonstrate this new way to thermally control the switching of a magnetic film and how it can be used with a GMR read out layer.

\section{Experimental Details}

The complete device structure is shown in Fig 1. All films were deposited on thermally oxidized Si substrates using magnetron sputtering at a base pressure better than $5\cdot10^{-8}$ Torr. The bottom Ni$_{80}$Fe$_{20}$ (NiFe) film was used as an underlayer to promote the growth of AFM Ir$_{20}$Mn$_{80}$ (IrMn)  \cite{ref:Berkowitz}. The IrMn was used to exchange bias a Co$_{90}$Fe$_{10}$ (CoFe) reference layer in a CoFe/Cu/NiFe GMR read out layer. A Ni-Cu alloy was cosputtered from Ni and Cu targets such that the Curie temperature of the alloy was just above room temperature. On top of the alloy an exchange biased NiFe layer was deposited. Finally a Ta capping layer was added to prevent oxidation. The argon pressure during sputtering was kept at 3 mTorr. To induce exchange bias at the IrMn/FM interfaces the whole deposition process was performed in a magnetic field of 350 Oe.

\begin{figure}[!t]
\centering
\includegraphics[width=3in]{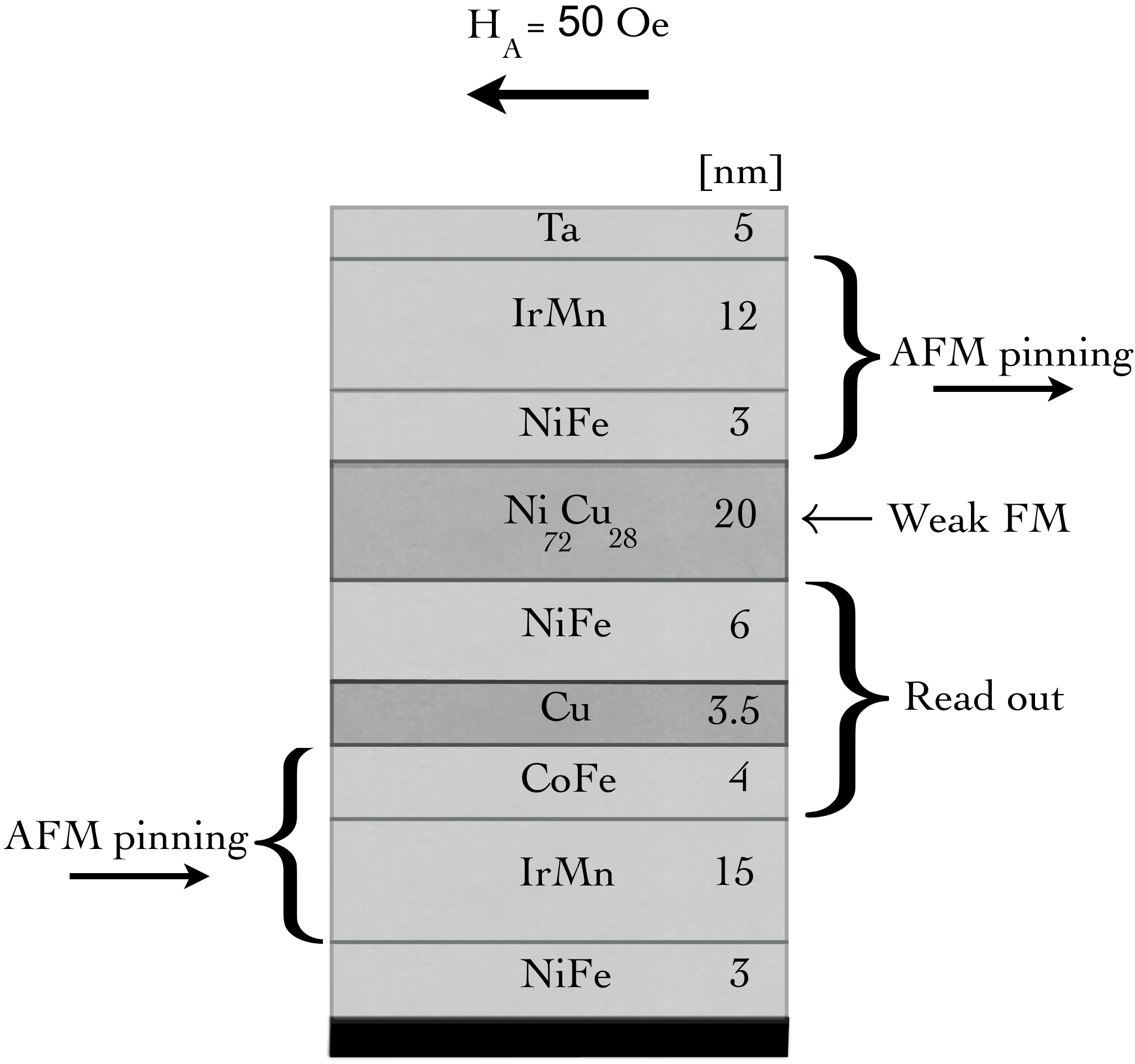}
\caption{The complete device structure deposited on a thermally oxidized Si substrate by magnetron sputtering. During measurements the device is placed in an external magnetic field, $H_{A}=50$ Oe, directed opposite to the AFM pinning direction.}
\label{fig_stack}
\end{figure}

To characterize the samples the temperature was raised using a thin-film heater, during which time current in plane (CIP) GMR measurements were performed. The temperature was controlled through a feed back loop using a type-T thermocouple in close contact with the sample.

Using photolithography the stacks were patterned into strips 50 $\mu$m wide and 1 mm long and then bonded at the edges with aluminum wires. During measurements the current was swept between $\pm$50 mA corresponding to a maximum current density of $10^{6}$A/cm$^{2}$. The resistance of the strips were obtained using four point measurements. While sweeping the current a constant field of 50 Oe was applied directed opposite to the AFM pinning direction.

\section{Results and Discussion}
At temperatures higher than room temperature the Ni-Cu alloy becomes paramagnetic and the magnetic moment of the NiFe read out layer is free to rotate in the external field (see Fig. 1). At temperatures closer to room temperature the alloy is ferromagnetic and effectively exchange couples to the read out layer. Because of this exchange coupling the magnetic moment of the NiFe read out layer will turn and point in the opposite direction to the external field. By varying the temperature, either externally or internally, we can thus control the direction of the NiFe magnetic moment in the read out tri-layer.

\subsection{Antiferromagnetic Pinning}
IrMn was chosen for the AFM pinning because of its high blocking temperature and thermal stability \cite{ref:Samant}. To reach blocking temperatures above 150$^{\circ}$C the films must be 10 nm or thicker \cite{ref:Devasahayam}. However, as the thickness increases the exchange bias decreases. It is important that the bottom CoFe reference layer stays exchange biased when the Ni-Cu alloy goes through the FM to paramagnetic phase transition. For all samples measured the bottom AFM had a thickness of 15 nm and the top AFM a thickness of 12 nm. 

\begin{figure}[!t]
\centering
\includegraphics[width=3in]{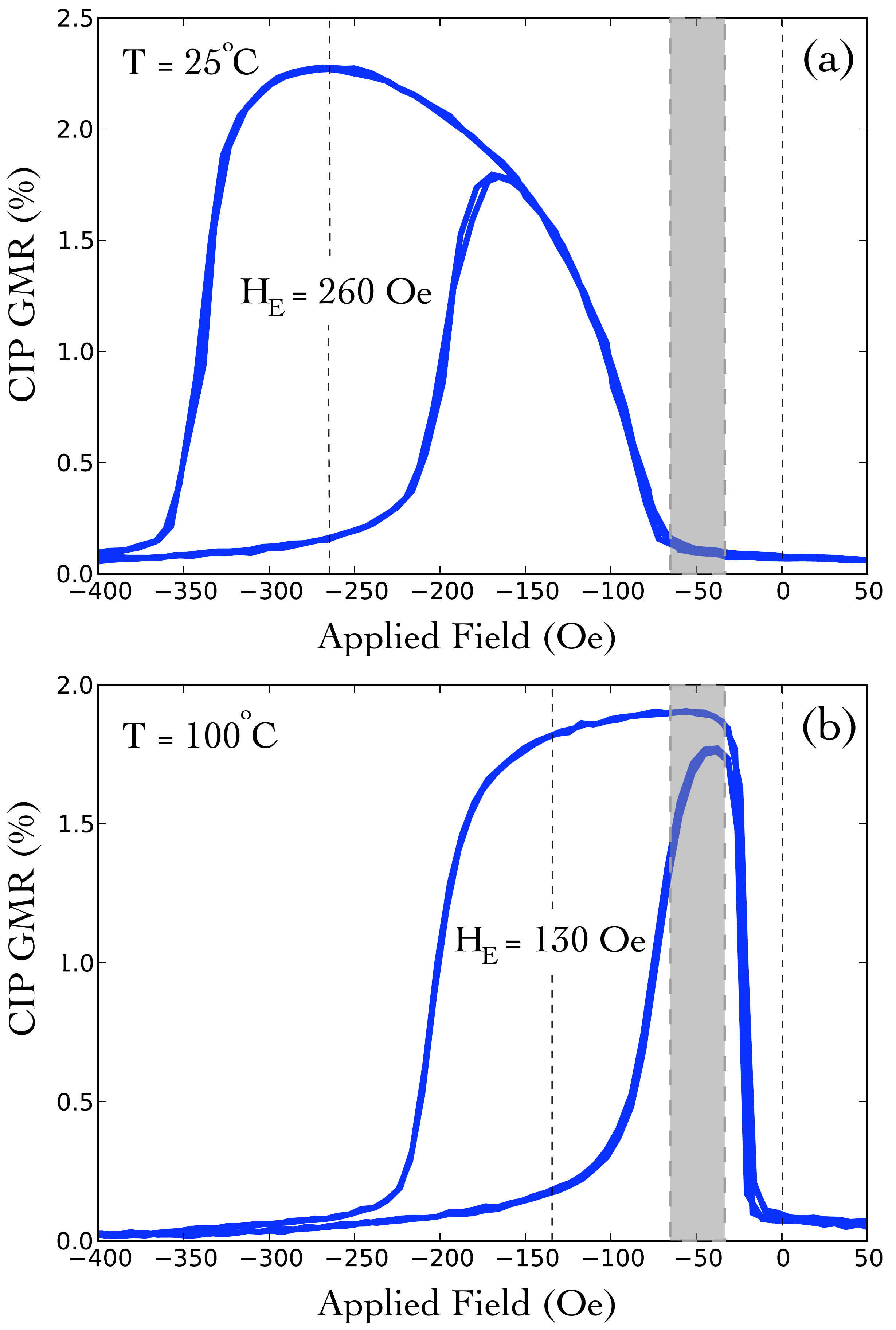}
\caption{Current in plane giant magnetoresistance versus applied field for a sample at (a) room temperature and (b) 100$^{\circ}$C heated with an external heater. The grayed out area indicates the region in which the external field should be applied during device operation.}
\label{fig_heater}
\end{figure}

Fig. 2 shows the GMR versus applied magnetic field for a typical sample (a) at room temperature and (b) at 100$^{\circ}$C. The temperature was controlled with an external heater. At room temperature the shift of the hysteresis loop, $H_{E}$, along the magnetic field axis for the CoFe reference layer is 260 Oe and decreases with increasing temperature.  At 100$^{\circ}$C $H_{E}$ has decreased to 130 Oe but is still well separated from the NiFe read out layer which has a loop shift of 20 Oe. By cycling the temperature to look for changes in $H_{E}$ after each completed cycle, it was concluded that neither of the AFM layers reach their blocking temperature. This procedure was repeated for all samples before patterning.

From the data in Fig. 2 it can be seen that the external magnetic field, $H_{A}$, should be in the range 70 to 30 Oe (grey area) applied opposite to the AFM pinning direction. If $H_{A}$ is too large the NiFe moment switches already at room temperature. If $H_{A}$ is too small nothing will happen to the read out layer as the NiFe layer is coupled to the CoFe reference layer --- indicated by the 20 Oe loop shift at 100$^{\circ}$C. This is probably due to N\'eel coupling through the 3.5 nm thick Cu spacer \cite{ref:Kools}. However, any residual exchange coupling through the Ni-Cu spacer cannot be completely ruled out \cite{ref:Hernando}.

\subsection{Ni-Cu Alloying}
The Curie temperature of the Ni-Cu alloy decreases linearly with decreased Ni concentration and reaches 0 K at a concentration of 44 at.\% Ni \cite{ref:Hicks}. In order to get the Curie temperature just above room temperature the concentration of Ni should be between 70-80  at.\% \cite{ref:Dutta}-\cite{ref:Sousa}. By varying the deposition rates of Ni and Cu during co-sputtering different Ni-Cu alloys were obtained. To find the optimal alloy composition the FM to paramagnetic phase transition of the alloys were characterized using a magnetometer with a built in heater. At 72 at.\% Ni  the alloy is FM at room temperature and can be used to couple two FM films. At 100$^{\circ}$C the phase transition is completed and the two FM films decouple. In order to completely separate two FM films at high temperatures the thickness of the Ni-Cu alloy has to be in the range 20-30 nm. A possible explanation for this is that the alloy is not completely homogenous after co-sputtering at room temperature but contain regions with higher Curie temperature. If the Ni-Cu spacer is very thin these regions could extend to the alloy interfaces and couple the two FM films. An extreme case would be if the alloy contained clusters of pure Ni with a Curie temperature of 354$^{\circ}$C. Another possible explanation is that the two FM films are coupled by exchange interactions through the spacer, even when the spacer is in the paramagnetic phase. It has been indicated that exchange can propagate through paramagnetic regions on length scales of several nanometers \cite{ref:Hernando}.

\subsection{Thermoelectrically Controlled Switching}

\begin{figure}[!t]
\centering
\includegraphics[width=3in]{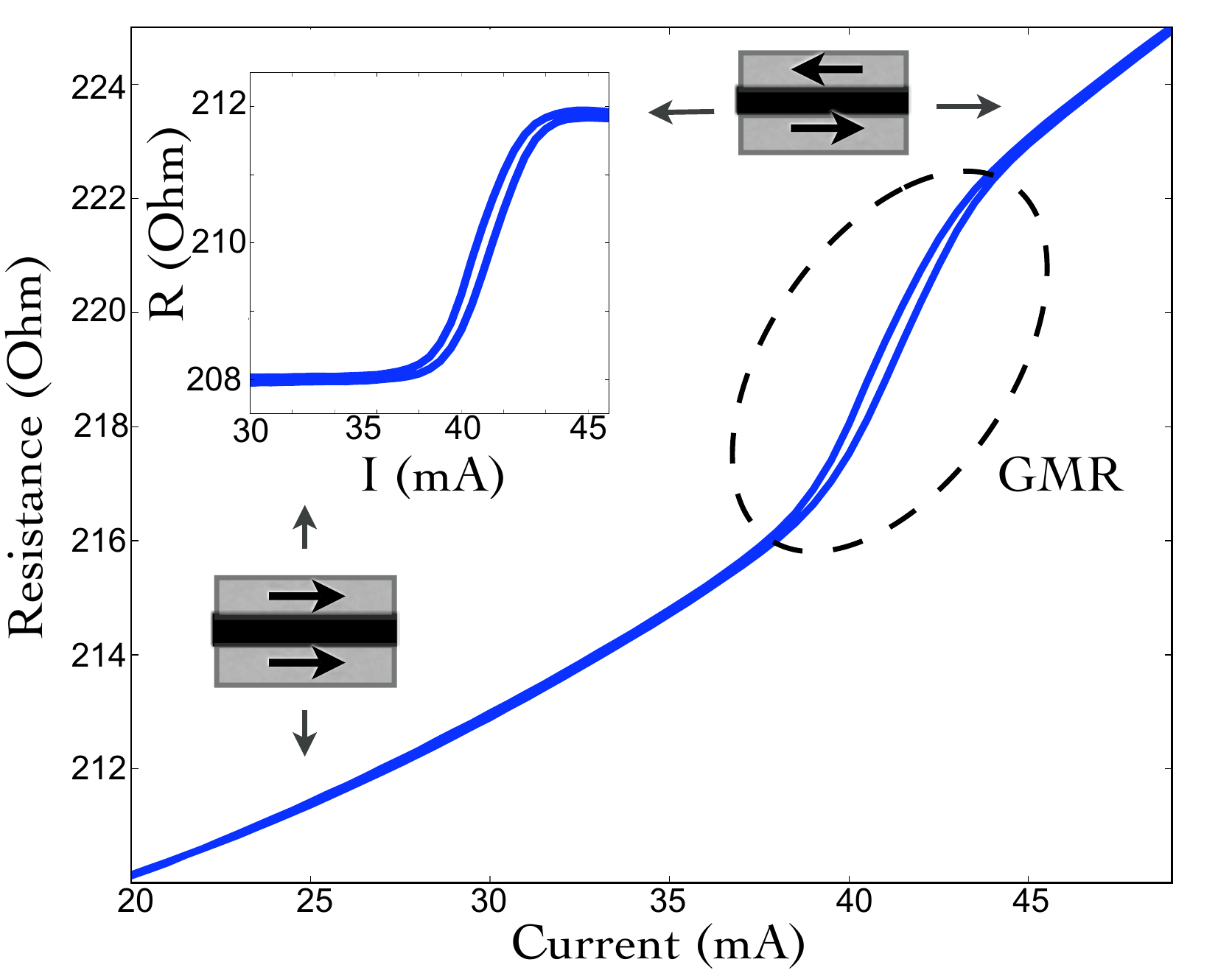}
\caption{Resistance versus current in plane for a 50 $\mu$m wide and 1 mm long strip. The circled part shows the region where the read out layer goes from parallel to antiparallel. Inset shows only the GMR transition with the thermal background subtracted.}
\label{fig_IV}
\end{figure}

Fig. 3 shows how the switching of the NiFe read out layer is controlled by current flowing through a 50 $\mu$m wide and 1 mm long strip. The intrinsic heating caused by the current is used to increase the temperature of the strip. At first the resistance changes linearly with temperature, which scales with the square of the heating current. However, at 38 mA the NiFe read out layer looses the exchange coupling and starts to rotate in the external magnetic field. Note that the read out layer is situated close to the center of the stack, and therefore not affected by the self-field of the current through the strip. The nonlinear resistance change continues until a current of 44 mA is reached. At higher currents the two FM films in the read out layer are completely antiparallel and the resistance increases linearly with temperature again. When the current is decreased the sample cools down and the exchange coupling through the Ni-Cu alloy comes back. No degradation of this behavior when cycling the current can be observed. The inset to Fig. 3 shows the change in resistance related to the FM to paramagnetic phase transition only. The thermal background, which scales with the square of the current, has been subtracted.

The change in current required to rotate the magnetization 180$^{\circ}$ depends on a number of factors. First there is the temperature change, $\Delta T$, required for the transition between the parallel and the antiparallel state. By looking at CIP GMR versus field curves for different temperatures (similar to Fig. 2) it can be concluded that $\Delta T$ for the sample in Fig. 3 is 35$^{\circ}$C. Less current would be required if a FM alloy with a sharper phase transition, $\Delta T<35^{\circ}$C, could be found. This would result in lower power consumption and faster switching.

The small hysteresis seen i Fig. 3 is caused by the free layer coercivity. If the read out layer produced the same GMR signal but had a film with higher coercivity, this hysteresis would be more significant and a larger change in current would be required to switch the magnetization back and forth.

Another way to decrease the required current is to increase the change in resistance between the parallel and antiparallel state. Then less current would be needed in order to achieve the same $\Delta T$. In this experiment the device is operated with current flowing in the plane of the film. However, if the device would be operated with current flowing perpendicular to the plane an increase of the magnetoresistance signal with roughly a factor of two is expected \cite{ref:Bass}. Different materials in the read out layer could also give a better signal. In a test to increase the signal the NiFe read out layer was replaced by a CoFe/NiFe/CoFe tri-layer. Fig. 4 shows the resulting minor loop at 100$^{\circ}$C compared to a sample with only a NiFe layer. As expected, the signal increases to almost 3\% but due to the inherent crystalline anisotropy of CoFe the coercivity also increases.

\begin{figure}[!t]
\centering
\includegraphics[width=3in]{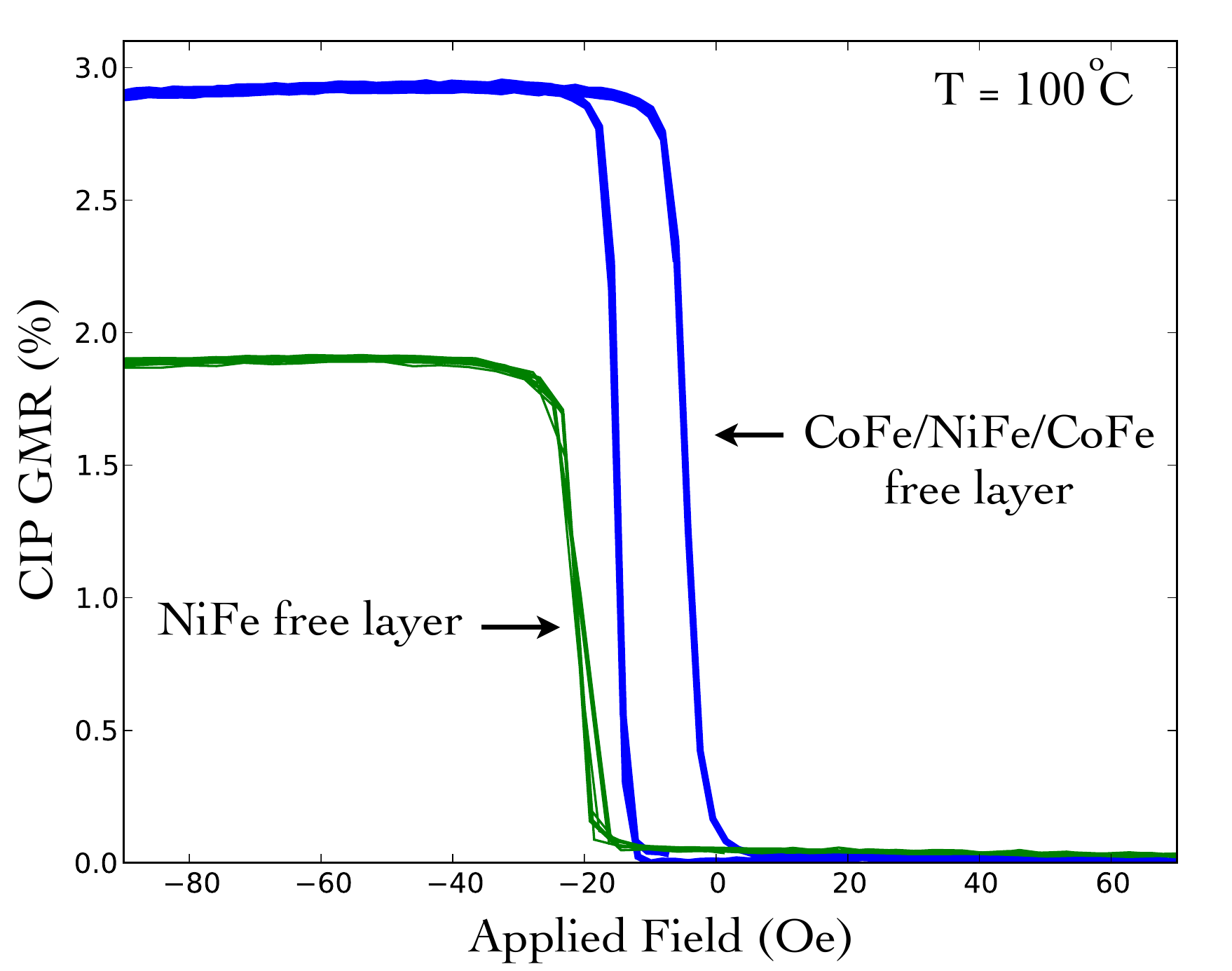}
\caption{Current in plane giant magnetoresistance versus applied field for a sample with a CoFe/NiFe/CoFe tri-layer as the free layer together with a sample having only a NiFe free layer to show the difference in signal and coercivity. The measurement was done at a temperature of 100$^{\circ}$C. }
\label{fig_read_out}
\end{figure}

The best way to increase the signal and decrease the required current would be to use tunneling magnetoresistance (TMR) instead of GMR for read out \cite{ref:Julliere}. A signal increase with a factor of 10--100 is expected \cite{ref:Yuasa}-\cite{ref:Parkin}. However, because of the rapid drop in TMR with applied voltage the resistance area product of the tunnel barrier must be relatively low in order to achieve the power densities required for heating. For heating a magnetic stack containing a tunnel barrier typical current densities of $10^{6}$A/cm$^{2}$ are used \cite{ref:Prejbeanu}. Assuming the device operates at 300 mV the required resistance area product would be 30 $\Omega \mu $m$^{2}$.

Before demonstrating the above thermoelectrically controlled switching in a device, the switching speed should be considered. For large samples the switching speed is limited by the thermal time constant. In order to measure the time constant the sample was heated to a temperature above room temperature. The heater was then turned off and the resistance change versus time measured. As expected for a 50 $\mu$m wide and 1 mm long strip, a very large time constant of 30s was recorded. If fast switching is required the size of the device would have to be decreased. For submicrometer sized junctions typical time constants are in the order of nanoseconds \cite{ref:Kerekes}. Since the device is operated by changes in current any series inductance could also limit the switching speed and has to be taken into account. 

\section{Conclusion}
We have successfully demonstrated thermoelectrically controlled switching of a magnetic film element. The switching is well controlled by external or internal heating. In order to completely reverse the magnetization of the film a temperature increase of 35$^{\circ}$C is needed. No degradation upon thermal cycling is observed. For optimal performance the device should be small in size, have a read out layer with large magnetoresistance and minimal free layer coercivity. This new way to thermoelectrically control magnetic switching in magnetic multilayers is promising for applications such as high frequency oscillators \cite{ref:Kadigrobov}.

\section*{Acknowledgment}
This work was supported by EU-FP7-FET-STELE.

\end{document}